\documentclass[submission,copyright,creativecommons]{eptcs}
\providecommand{\event}{ICLP 2025} 

\usepackage{iftex}

\usepackage{times}
\usepackage{soul}
\usepackage{url}
\usepackage{hyperref}
\usepackage{graphicx}
\usepackage{amsmath}
\usepackage{amsthm}
\usepackage{booktabs}
\usepackage{algorithm}
\usepackage{algorithmic}
\usepackage{bm}
\usepackage{bbm}
\usepackage{amssymb}
\usepackage[english]{babel}
\usepackage[small]{caption}
\usepackage{xcolor}
\usepackage{listings}
\usepackage{pdfpages}
\usepackage{textcomp}

\makeatletter
\newcounter{pagecount}
\newcommand{\limitpages}[1]{
    \setcounter{pagecount}{0}%
    \gdef\maxpages{#1}%
    \ifx\latex@outputpage\@undefined\relax%
        \global\let\latex@outputpage\@outputpage%
    \fi%
    \gdef\@outputpage{%
        \addtocounter{pagecount}{1}%
        \ifnum\value{pagecount}>\maxpages\relax%
        \else%
            \latex@outputpage%
        \fi%
    }%
}
\makeatother

\lstdefinestyle{gringo}{
    language=Prolog, 
    basicstyle=\ttfamily,
    keywordstyle=\color{teal},
    commentstyle=\color{gray},
    stringstyle=\color{red},
    numbers=left,
    numberstyle=\tiny\color{gray},
    stepnumber=1,
    numbersep=5pt,
    frame=single,
    tabsize=2,
    breaklines=true,
    morekeywords={reached, goal_gState, gState, trans, obsi, doi, dist, w, action, cell, block, goal, goal1, goaln, goali, move, aState, near, goali_aState, avai_action, isAction, cost, nextTo, lastmin, traffic}, 
    sensitive=true,
    literate={not}{{\textcolor{purple}{not}}}1
}

\newtheorem{theorem}{Theorem}

\newtheorem{proposition}{Proposition}
\newtheorem{definition}{Definition}

\ifpdf
  \usepackage{underscore}         
  \usepackage[T1]{fontenc}        
\else
  \usepackage{breakurl}           
\fi

\title{Computing Universal Plans for Partially Observable Multi-Agent Routing Using Answer Set Programming}
\author{
Fengming Zhu \qquad\qquad Fangzhen Lin
\institute{
Department of Computer Science and Engineering\\
Hong Kong University of Science and Technology\\
Hong Kong SAR, China}
\email{fzhuae@connect.ust.hk \quad flin@cs.ust.hk}
}
\def\titlerunning{Universal Planning for PO-MAR using ASP}
\def\authorrunning{Fengming Zhu \& Fangzhen Lin}
\begin{document}
\maketitle


\begin{abstract}
Multi-agent routing problems have gained significant attention recently due to their wide range of industrial applications, ranging from logistics warehouse automation to indoor service robots.
Conventionally, they are modeled as classical planning problems.
In this paper, we argue that it can be beneficial to formulate them as universal planning problems, particularly when the agents are autonomous entities and may encounter unforeseen situations.
We therefore propose \textit{universal plans}, also known as \textit{policies}, as the solution concept, and implement a system based on \textit{Answer Set Programming} (ASP) to compute them.
Given an arbitrary two-dimensional map and a profile of goals for a group of \textit{partially observable} agents, the system translates the problem configuration into \textit{logic programs} and finds a feasible universal plan for each agent, mapping its observations to actions while ensuring that there are no collisions with other agents.
We use the system to conduct experiments and obtain findings regarding the types of goal profiles and environments that lead to feasible policies, as well as how feasibility may depend on the agents' sensors.
We also demonstrate how users can customize action preferences to compute more efficient policies, even (near-)optimal ones. The code is available at \url{https://github.com/Fernadoo/MAPF_ASP}.
\end{abstract}

\section{Introduction}
Consider a situation
where a group of robots are programmed to navigate themselves between pre-specified task points.
The robots are expected to reach their task points as quickly as possible while avoiding collisions with one another.
This problem is commonly encountered in automated warehouses and is typically formulated as a \textit{Multi-Agent Path Finding} problem (MAPF), which is mostly solved using centralized algorithms~\cite{stern2019multi-overview,stern2019multi}.
Although optimally solving MAPF problems is in principle \textit{NP-hard}~\cite{yu2013structure},  in practice there are planners that can efficiently find high-quality solutions even for large-scale instances~\cite{sharon2015conflict,li2021eecbs,okumura2022priority,okumura2023lacam}.

However, beyond the logistics industry, the use of such mobile robots in household or office scenarios is also of significant commercial value.
The latter presents at least the following distinct features: 1) there is typically no central server computing plans and communicating with all the robots in real time; 2) each robot is usually equipped with some sensors and computing resources, although the capabilities may be limited to maintain cost efficiency; 3) the domain can be relatively small compared to industrial warehouses.
These inherently different features may necessitate a potentially different solution concept.
More specifically, a feasible solution in an MAPF problem is defined as \textit{a set of collision-free paths} (i.e., sequences of actions), one for each agent.
If any contingency occurs, such as one of the robots being perturbed from its planned path or simply stopping accidentally, the entire joint plan must be re-computed in a centralized manner.
As mentioned, unlike the situation in a warehouse, home service robots typically do not have access to a central controller during real-time execution.
Instead, users may use a smartphone APP to synchronize their robots at the beginning of the deployment, and leave them for their own fully autonomous routing for a quite long period.
Therefore, we propose to adopt the solution concept as \textit{a set of functions}, each mapping an individual agent's sensory readings (also known as observations or local states\footnote{We will use these three terms interchangeably. Note that here local states do not refer to the states in any finite automaton.}) to their actions.
These functions are referred to as universal plans in planning~\cite{schoppers1987universal}, strategies in extensive-form games, and policies in Markov decision processes (MDPs)\footnote{We will use the terms universal plans and policies interchangeably.}.


\begin{figure}[!ht]
    \centering
	\includegraphics[scale = 0.5]{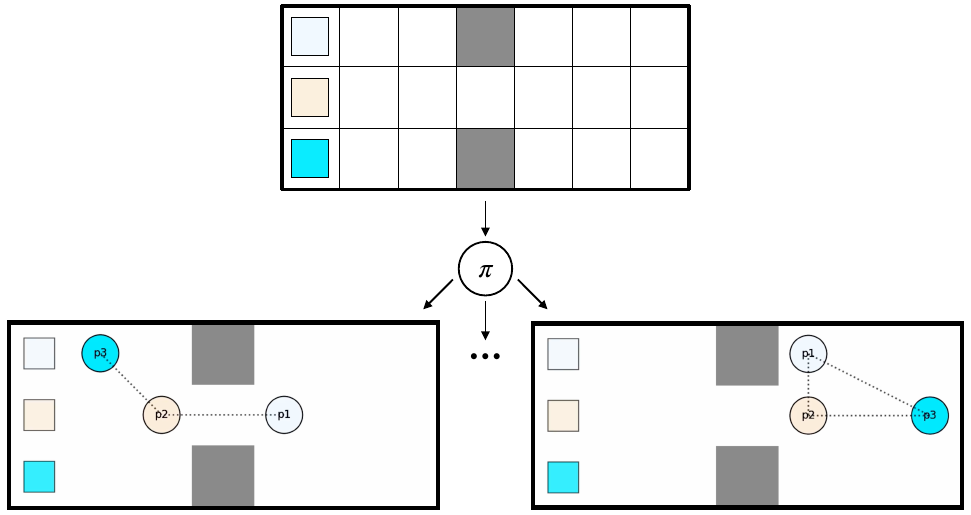}       
    \caption{An example (two chambers connected by a tight hallway)
    that shows how our proposed policy-like solution works.
    Once computed (taking 167.35s for searching),
    it will be applicable for all $19\times 18\times 17 = 5814$ global states.
     Colored squares are the goals assigned to
    sphere-shaped agents in corresponding colors. When two agents are connected by a dotted line, it means they can detect each other.}
    \label{fig:policy}
\end{figure}

In this paper, we present a system based on \textit{Answer Set Programming} (ASP) for computing universal plans.
We assume multiple agents are moving on a 2D grid map with potential obstacles.
Each agent is aware of the map layout and has a designated goal location, but does not know the goals or current locations of the other agents unless they are within its field of view (FoV).
Moreover, none of the agents are equipped with communication modules for message sharing.
Essentially, our system takes in a problem instance specified by a given map layout and a group of partially observable agents, translates it into ASP encodings, and computes a decentralized policy for each agent.
In principle, once a policy profile is found,
it works no matter where the agents are placed initially and how they get perturbed during routing.
In other words, it can be seen as a paradigm of \textit{centralized search with decentralized execution}.
A graphical illustration is provided in Figure \ref{fig:policy}.
With the help of this system, we conduct the following studies, presenting a spectrum ranging from feasibility to optimality.


\textbf{Feasibility}.
A policy profile is feasible if all agents can achieve their goals without ever colliding with others.
We evaluate the efficiency of our system and
report the time cost for computing feasible policies for various configurations.
Sometimes there may not exist any feasible policy, and we therefore design specific experiments to investigate how the existence of feasible policies depends on the agents' sensory capabilities.

\textbf{Action heuristics}.
To take one step further,
one may resort to certain heuristics the improve the quality of the solution policy.
Our system provides two ways for the users to specify action heuristics.
The first type of heuristic is defined as each individual agent's preferences over actions, e.g., an agent may prefer actions that bring her closer to the goal, regardless of her observations.
The second type of heuristic can be interpreted as a set of traffic rules that require agents to adopt a shared protocol when they encounter one another, which is represented as common knowledge among all agents.
These action heuristics rule out undesirable policies, and can therefore be an effective strategy for pruning the search space.

\textbf{Anytime optimization}.
Modern ASP solvers enable us to compute optimized solutions in an \textit{anytime} fashion.
The longer the program runs, the better the returned solution will be.
However, given the high complexity of the overall problem, our experiments show that it is computationally more practical to implement heuristic functions through action restrictions to compute better policies, rather than optimizing from scratch by exhaustive search.

The rest of the paper is organized as follows.
Section \ref{sec:related_work} lists a number of related domains.
After that,
Section \ref{sec:preli} introduces some mathematical notations to illustrate the problem.
Next, Section \ref{sec:encoding} elaborates on the formal encodings to address the problem.
Section \ref{sec:case_study} presents several experiments and findings.
Finally, we conclude the paper by discussing potential applications and future directions.

\section{Related Work}
\label{sec:related_work}
We list a few related areas that inspire our work,
but still, we need to distinguish our work from those.

\textbf{Compilation-based methods for MAPF}.
For conventional MAPF problems, apart from leveraging
best-first search techniques~\cite{sharon2015conflict,li2021eecbs},
researchers also investigate methods that compile MAPF instances into formal languages
and solve them by declarative programming.
Success has been witnessed in formalizations such as
SAT~\cite{surynek2016efficient},
CSP~\cite{wang2019new}, and
SMT~\cite{surynek2019unifying}.
In particular, we note that one can also formulate such problems in similar or variant settings in nonmonotonic formalisms like Answer Set Programming (ASP)~\cite{erdem2013general,bogatarkan2019declarative,gomez2020solving,bogatarkan2020explanation}.
Nevertheless, the solution concept in these settings is still a sequence of joint actions for a group of agents from their initial positions to goal positions.
 Our work in this paper aims to compute a universal plan, i.e. a profile of decentralized individual policies, 
 each of which maps an agent's local states (defined under partial observability) to available actions. 
 The policy profile by definition should work for any possible situation.


\textbf{Universal planning}.
Another area that closely aligns with ours is known as \textit{universal planning} in single-agent domains.
Schoppers~\cite{schoppers1987universal}
first motivated the investigation of universal plans
by an example of the well-known \textit{Blocks World},
where a conventional planner may be bothered
by a naughty baby who would occasionally smash down the built blocks,
and therefore, may need repeated replanning.
The idea of universal planning in fact quite overlaps with \textit{contingent planning}
or planning under partial observability~\cite{bonet2011planning,zanuttini2020knowledge}. 
Work from those domains also desires policies or conditional plans as the appropriate solution concept.
Recent studies, such as~\cite{zanuttini2020knowledge},
have thoroughly investigated the idea of expressing policies by knowledge-based programs,
where branching conditions are represented as epistemic formulas~\cite{sep-logic-epistemic},
even in partially observable domains.
This solution concept can also be implemented using ASP in applications related to cognitive robotics~\cite{yalciner2017hybrid}.
In a comprehensive study, Son et al.~\cite{tran2023answer} have surveyed the maturity and drawbacks of utilizing ASP for planning problems.
Therefore, our work can be seen as a multi-agent extension, where each agent is required to execute the distributed plan from its own side.

\textbf{Dec-POMDPs and MARL}.
The most related and potentially the most popular domain is \textit{Multi-Agent Reinforcement Learning} (MARL), particularly for cooperative tasks.
The underlying model is typically framed as a \textit{Decentralized Partially Observable Markov Decision Process} (Dec-POMDP), which is a Markov chain of states where agents share the same utility function but are not fully aware of their current states~\cite{kaelbling1998planning,littman2009tutorial,oliehoek2016concise}.
To solve Dec-POMDPs, dynamic programming~\cite{hansen2004dynamic} based on Bellman's principle is typically employed, but it suffers from notorious complexity, as it is generally undecidable to optimally solve infinite horizon Dec-POMDPs~\cite{madani1999undecidability}.
In contrast, our work approaches the problem from the perspective of satisfiability, utilizing efficient constraint propagation and effective heuristics.
As a result, our system is capable of solving larger problem instances.
There is also a vast body of literature that leverages reinforcement learning techniques to find approximate solutions~\cite{lowe2017multi,foerster2018counterfactual,rashid2020monotonic,yu2022the}.
However, learning-based frameworks cannot always guarantee feasible solutions, as they primarily optimize cumulative rewards, which sometimes cannot well capture hard constraints in planning problems.
Our system can, to some extent, be viewed as a program that automatically generates "safety rules", but we leave the exploration of how to integrate a logic program into a learning framework for future work.



\section{Problem}
\label{sec:preli}
We consider a group of agents with partial observability moving along a 4-neighbor grid map.
Formally, a \textit{configuration} is a 5-tuple $\langle G, N, A, \{r_i\}_{i\in N}, \{g_i\}_{i\in N} \rangle$ defined as follows,
\begin{itemize}
	\item The 4-neighbor grid map $G = (V, E)$ is an undirected graph, where $V$ is the set of vertices that an agent can be placed at, and $E$ is the set of unit-cost edges.
	\item $N$ is a set of $n$ agents. Each agent is associated with five unit-cost actions $A =$ \{\texttt{up}, \texttt{down}, \texttt{left}, \texttt{right}, \texttt{stop}\} with their usual meanings. Each agent $i$ is equipped with a sensor of range $r_i$, enabling it to detect another agent at locations $D_i^v = \{u\in V: d(u,v)\leq r_i\}$ given its own location $v \in V$, where $d: V\times V \mapsto \mathbbm{R}$ denotes some distance measure. We also term the detected area as each agent's \textit{field of view} (FoV).
	\item $\{g_i\}_{i\in N}$ is a goal profile, where $g_i$ is the goal designated to agent $i$.
\end{itemize}
The above ingredients consequently define the following notions of \textit{global states} and \textit{local states},
\begin{itemize}
	\item The set of global states $S \triangleq V^N$ consists of all possible joint-location tuples of all agents.
	\item The set of local states $O_i \triangleq \{(v, u): u \in D_i^v \lor u = \texttt{empty} \}_{v \in V}$ for each agent $i$ consists of its own locations and observations of the others, including the case when it sees no one within its sensor.
\end{itemize}

A \textit{collision} happens
when two agents simultaneously move to the same vertex (i.e., vertex conflict)
or traverse the same edge (i.e., edge conflict)~\cite{stern2019multi}.
Mathematically, let $pos_i^t$ denote the vertex
where agent $i$ is positioned at time step $t$,
then a collision happens between agent $i$ and $j$ if
\begin{itemize}
    \item $pos_i^t \in V \land pos_j^t \in V \land pos_i^t = pos_j^t$, or
    \item $(pos_i^t, pos_j^t) \in E \land (pos_i^{t+1}, pos_j^{t+1}) \in E
            \land (pos_i^t, pos_j^t) = (pos_j^{t+1}, pos_i^{t+1})$.
\end{itemize}

An \textit{instantiation} is to assign an initial profile $\{\alpha_i\}_{i \in N} \in V^N$ to agents,
where $\alpha_i$ is the initial position of agent~$i$ and $\alpha_i \neq \alpha_j, \forall i \neq j$.
This means that no two agents can be placed at the same initial position.
We desire a deterministic \textit{policy} $\pi_i: O_i \mapsto A$ for each agent~$i$, which is a mapping from its local states to actions.
Note that an intermediate perturbation (or replacement) of agents' locations during routing is equivalent to instantiating them in different initial positions and let them re-route.
Given a configuration, a \textit{policy profile} $\{\pi_i\}_{i \in N}$ is \textit{feasible} if the following three conditions are satisfied for all possible instantiations,
\begin{enumerate}
	\item No collision happens between any pair of agents.
	\item Once an agent has reached her goal,
	she will stop and serve as a fixed obstacle.
	\item All agents can successfully reach their goals within finite steps.
\end{enumerate}
A configuration is \textit{solvable} if there exists a feasible policy profile for this configuration.

\section{Formal Encoding}
\label{sec:encoding}
In this section, we formalize the problem
via the language of \textit{Answer Set Programming} (ASP).
We address the problem by first finding feasible solutions
and then incorporating optimization procedures.
Given that ASP is a prevalent formalism nowadays, we will omit its introduction here.
Readers interested in the detailed syntax and stable model semantics are encouraged to refer to \cite{vladimir2008answer, marek1999stable, niemela1999logic, gelfond1988stable}.
An advantage of encoding this problem into logic programs is that it allows us to easily incorporate human-readable rules to study the behavior of the agents and potentially enhance the quality of the policy.

Some readers may find the enumeration of the encoded rules tedious. Therefore, we have postponed most of the encodings of the facts and domain definitions that are already clear from the previous section to Appendix~\ref{apd:prob_encoding} for better presentation.
Here, we present several key rules that reveal the nature of this problem.
To ensure that the policy profile works for all possible situations,
we inductively define the reachability of each global state given the policy (\texttt{reached/1}).
Basically, a global state (\texttt{gState/1}) is trivially reachable to itself if all agents have already arrived at their goals (\texttt{goal_gState/1}).
Inductively, any global state is reachable to the goal state
if it can, according to the policy, transit to a successor state that is reachable to goal state,
where the transition (\texttt{trans/3}) is made by a collection of all agents' actions (\texttt{doi/2}) based on their own observed local states (\texttt{obsi/2}). 
\begin{lstlisting}[style=gringo]
reached(S):- goal_gState(S).
reached(S1):- gState(S1), trans(S1,(A1,...,An),S2), reached(S2),
    obsi(Si,ASi), doi(ASi,Ai). % observe and act, for all i
:- gState(S),  not  reached(S). % rule out infeasible policies
\end{lstlisting}

\begin{theorem}[Correctness]
This ASP implementation (with the complete version attached in Appendix~\ref{apd:prob_encoding}) is both sound and complete.
More specifically, 1) given an answer set of a logic program described above,
the policy profile encoded in the answer set by the \verb-doi/2- atoms will be a feasible one;
2) conversely, one can construct an answer set of the above logic program from a feasible policy profile. 	
\end{theorem}

The justification is quite straightforward.
For one, we have directly encoded those three conditions that define a feasible policy profile.
More specifically, 1) in the rule that defines valid global states (\texttt{gState/1}), it prohibits any pair of agents from sharing the same location; 2) in the rule that defines valid transitions (\texttt{trans/3}), it eliminates any immediate location swaps; and 3) The above reachability rule ensures that every global state can reach the goal state within a finite number of transitions (\texttt{reached/1}).
Note that the reachable states eventually form a spanning tree rooted at the goal state, with each internal node serving as the root for the states within its subtree.

For optimization, there are several evaluation criteria used in conventional MAPF problems.
However, none of them is suitable for our settings,
as we require a policy to work in all situations.
Thus, we adopt a criterion as minimizing \textit{sum-of-makespan},
which is the sum of the makespans of the execution of the policy under all possible instantiations.
Mathematically,
$$
\textit{sum-of-makespan} (\{\pi\}_{i \in N})=
\sum_{\{\alpha_i\}_{i \in N} \in V^N: \alpha_i \neq \alpha_j} \textit{makespan}(\{\alpha_i\}_{i \in N}, \{\pi\}_{i \in N}).
$$
where the metric \textit{makespan} is defined as the number of transitions
made from the initial global state to reach the goal global state.
Therefore, we can also define \textit{makespan} inductively as follows, denoted as the \texttt{dist/2} predicate.
A transition between two global states will incur a unit cost.
\begin{lstlisting}[style=gringo]
w(0..UPPERBOUND). % pre-defined distance range
dist(S,0):- goal_gState(S).
dist(S1,C):-  not  goal_gState(S1), trans(S1,(A1,...,An),S2),
    reached(S1), dist(S2,C-1), w(C),
    obsi(Si,ASi), doi(ASi,Ai). % for all i
#minimize {C, S: dist(S,C)}.
\end{lstlisting}

Lastly, we emphasize that the computed profile of policies is a collection of \textit{deterministic} and \textit{decentralized} ones,
meaning that once deployed on individual agents, each agent will be able to execute the policy independently,
and no further intervention from any central controller will be needed. 

\section{Experiments and Findings}
\label{sec:case_study}
In this section, we categorize our experiments and corresponding findings into three folds.

First, we aim at feasible policies.
For the readers' reference, we report the time cost to compute feasible policies for various configurations.
Besides, comprehensive experiments are designed to investigate
how feasibility depends on agents' sensors.
We leave the theoretical characterization for future work.

Second, we restrict the choices of agents' actions by certain heuristics to see if there exists a better yet feasible policy.
When the environment is sparse, once we have set appropriate action heuristics,
the system may already be efficient enough,
as agents normally do not need to sacrifice their own utilities too much due to potentially unnecessary coordinations with others.

Finally, we present two alternatives to optimize the policy.
We note that implementing heuristic functions through action restrictions is more effective for computing better policies than optimizing from scratch by exhaustive search.



\subsection{Feasibilities}
\subsubsection{Computational Time}
For a rough analysis of the time complexity, each agent is associated with
around $\Theta \bigl( M{\binom{K+n-1}{n-1}} \bigr)$ local states,
where $M$ denotes the number of all possible positions
which an agent can be placed at and $K$ denotes the number of neighboring cells that an agent can detect.
As mentioned, the formalized problem is also compatible with the Dec-POMDP formulation.
To demonstrate that the capability of our approach excels that of both classical Dec-POMDP solvers~\cite{bernstein2009policy, amato2010optimizing} and some up-to-date ones~\cite{kumar2016dual,koops2024approximate}, we list several benchmarks that are used to evaluate Dec-POMDP solvers in Table~\ref{tab:decpomdpbench}.
One can clearly see that the scales of our tested cases significantly exceed those of the Dec-POMDP benchmarks. For example, for a $4\times 4$ map with three agents, each equipped with sensors of range two (Grid4x4\_a3\_s2),
then each agent will be faced with around 2000 local states,
which is an intractable scale for Dec-POMDP solvers.

We report the time cost for computing a feasible policy in Table \ref{tab:comp_time}.
Programs are solved by \textit{clingo}\footnote{https://potassco.org/clingo/}~\cite{gebser2007clasp,gebser2011potassco,gebser2019multi}, and all experiments are done on Linux servers with Intel\textsuperscript{\textregistered} Xeon\textsuperscript{\textregistered} Silver 4210 CPU @ 2.20GHz.
\textit{Sensor r}
means that the agents are equipped with sensors of range $r$.\footnote{For the rest of this paper,
we will term \textit{sensor $r$} and sensors of range $r$ interchangeably.}
More specifically, following our previous notation, a sensor of range $r$ enables an agent to detect
any neighboring agent within an $L_\infty$ distance\footnote{For all the experiments, we adopt the $L_\infty$ metric as the distance measure, i.e. $dist(p, q) \doteq \max\{|p_x - q_x|, |p_y - q_y|\} \leq r$. One can of course implement any other options, e.g., Manhattan or Euclidian distance.} of no more than $r$.
It is observed that different goal locations subtly affect the computational time for a given map, therefore, the results are collected from a single goal profile for each configuration.
Note that, unlike learning-based systems, our system solves exact solutions that guarantee feasibility.
Once a feasible policy profile is computed, it can be distributed to each agent, and no replanning will be needed regardless of where the agents are positioned on the map.

\begin{table}[tb]
\small
\centering
\begin{tabular}{@{}lllll@{}}
\toprule
                & \#(agents) & \#(states) & \#(actions/agent) & \#(observations/agent) \\ \midrule
\textit{\textbf {Dec-POMDP benchmarks}}~\cite{bernstein2009policy, amato2010optimizing,kumar2016dual,koops2024approximate}                                              \\
Tiger           & 2          & 2          & 3                 & 2                      \\
Hotel           & 2          & 16         & 3                 & 4                      \\
Recycling       & 2          & 4          & 3                 & 2                      \\
Grid2x2         & 2          & 16         & 5                 & 2                      \\
Grid3x3         & 2          & 81         & 5                 & 9                      \\
\vspace{-3mm}
\\
\textit{\textbf {Our tested cases (selected)}}                                          \\
Grid3x3\_a2\_s1 & 2          & 72         & 5                 & 48                     \\
Grid6x6\_a2\_s3 & 2          & 1260       & 5                 & 896                    \\
Grid4x4\_a3\_s2 & 3          & 3360       & 5                 & 2196                   \\
Grid6x6\_a3\_s3 & 3          & 42840      & 5                 & 22568                  \\ \bottomrule
\end{tabular}
\caption{Benchmarks for evaluating Dec-POMDP solvers compared to those for our solver.
}
\label{tab:decpomdpbench}
\end{table}

\begin{table}[!t]
\small
\centering
\begin{tabular}{lrrrr}
\toprule
Map size    & 2 agents  & 3 agents  & 4 agents     & 5 agents \\
\midrule
\textit{\textbf {Sensor 1}} \\
$3\times3$  &  0.06s    & 3.23s     & 6.09min      & 8.76hr       \\
$4\times4$  &  0.29s    & 5.09min   & 28.40hr      & $\slash$     \\
$5\times5$  &  1.61s    & 40.24min  & $\slash$     & $\slash$     \\
$6\times6$  &  11.69s   & 6.02hr    & $\slash$     & $\slash$     \\
\vspace{-3mm}
\\
\textit{\textbf {Sensor 2}} \\
$3\times3$  & 0.06s     & 3.51s     & 8.86min      & 8.47hr      \\
$4\times4$  & 0.49s     & 2.22min   & 15.60hr      & $\slash$    \\
$5\times5$  & 2.30s     & 24.02min  & $\slash$     & $\slash$    \\
$6\times6$  & 7.50s     & 4.14hr    & $\slash$     & $\slash$    \\
\vspace{-3mm}
\\
\textit{\textbf {Sensor 3}} \\
$3\times3$  & 0.10s     & 3.37s     & 9.00min      & 8.92hr      \\
$4\times4$  & 0.39s     & 2.42min   & 18.69hr      & $\slash$    \\
$5\times5$  & 2.07s     & 26.71min  & $\slash$     & $\slash$    \\
$6\times6$  & 7.54s     & 4.46hr    & $\slash$     & $\slash$    \\
\bottomrule
\end{tabular}
\caption{Computational time for various maps, numbers of agents and sensors.
\textit{Sensor i}
means the agents are equipped with sensors of range $i$. We set the time limit to be 30 hours, and ``$\slash$'' means timeout.}
\label{tab:comp_time}
\end{table}

\subsubsection{Existence of Feasible Policies}
\label{subsubsec:existence}
In addition to scenarios where feasible policies are found, there are cases where no feasible policy exists.
Map layouts, sensor ranges, and goal placements all play a significant role in this.
We will first provide a necessary condition as follows.



\begin{definition}
	Given a map $G$, a goal profile $\{g_i\}_{i \in N}$ is said to be \textit{proper}, if for any $g_i$,
	there is always a path from any non-goal position $v\in V\backslash \{g_j\}_{j\neq i}$ to $g_i$,
	without passing through any of $\{g_j\}_{j\neq i}$.
	\label{def:proper_goal}
\end{definition}

\begin{proposition}
There exists a feasible policy only if the goal profile is proper.	
\end{proposition}

\begin{proof}
This \textit{necessary} condition can be easily justified.
Note that agents are not supposed to move any longer once they reach their own goals.
If the given goal profile is not proper, then if certain agents are placed
right at their goals, there is no chance for the rest to reach their goals.
\end{proof}

Therefore, the number of solvable cases (i.e., configurations under which a feasible policy exists) cannot exceed the number of proper goal profiles.
Besides, we also conduct experiments to find out which type of sensors will lead to feasible policies.
We start by testing some manually designed maps. It turns out that there may not exist any feasible policy for \textit{sensor 1}, while 
\textit{sensor 2} is capable enough of ensuring feasible policies for all possible proper goal profile at least in these tested cases.
For better presentation, we attach the detailed map layouts and statistics in Appendix~\ref{apd:manual}.
We additionally test the capability of \textit{sensor 1} versus \textit{sensor 2}
under randomly generated maps, with results reported in Table~\ref{tab:sensor_rand},
where $4\times4$ maps are associated with three agents
and $5\times 5$ maps are associated with two agents.
We report the number of goal profiles that allow feasible policies to exist,
the number of proper goal profiles,
as well as the total number of all randomly generated goal profiles.
As shown in Table \ref{tab:sensor_rand},
all profiles that cannot be solved by \textit{sensor 1} are solvable by \textit{sensor 2}
with a success rate of 100\%. 

\begin{table}[]
\small
\centering
\begin{tabular}{@{}lrr@{}}
\toprule
Map, \#agents, obstacle density              & sensor 1        & sensor 2                \\
 &       \#feasible/~\#proper/~\#total          &          \#feasible/~\#proper/~\#total      \\ 
\midrule
$4\times 4, 3\ agents, 6\%$          & $62.3/~~65.3/100$   & $\mathbf{65.3/~~65.3}/100$   \\
$4\times 4, 3\ agents, 12\%$         & $33.7/~~41.3/100$   & $\mathbf{41.3/~~41.3}/100$   \\
$4\times 4, 3\ agents, 18\%$         & $30.0/~~39.0/100$   & $\mathbf{39.0/~~39.0}/100$   \\
$5\times 5, 2\ agents, 10\%$         & $289.3/289.3/300$ & $289.3/289.3/300$                           \\
$5\times 5, 2\ agents, 20\%$         & $176.3/197.3/300$ & $\mathbf{197.3/197.3}/300$ \\
$5\times 5, 2\ agents, 30\%$         & $105.3/114.0/300$ & $\mathbf{114.0/114.0}/300$ \\ \bottomrule
\end{tabular}
\caption{For each setting, results are averaged over three random seeds. For each generated maps, we randomize 100 goal profiles for $4\times 4$ layouts
and 300 goal profiles for $5\times 5$ layouts.
$4\times 4$ layouts are tested for three agents and
$5\times 5$ layouts are tested for two agents.
We set three density levels of obstacles 6\%, 12\% and 18\% for $4\times 4$ layouts,
and 10\%, 20\% and 30\% for $5\times 5$ layouts.
Those configurations where \textit{sensor 2} is more capable than \textit{sensor 1}, the results are marked in bold fonts.
}
\label{tab:sensor_rand}
\end{table}

An applicable explanation is,
if agents are only equipped with \textit{sensor 1}, then they cannot detect each other when they are only two blocks away from each other, i.e., about to collide.
Consequently, the rules for eliminating vertex conflicts will in fact propagate more constraints during the solving process, potentially leading to the eventual unsatisfiability of the logic program.
More evidence will be shown later in Table~\ref{tab:restriction},
where maps in larger sizes and even action heuristics are considered.
As it will reveal,
even if we force agents to move greedily toward their goals
until when they have to coordinate to avoid collisions,
sensors of range two are still capable enough of working out feasible policies.


\subsection{Action Heuristics}
\label{subsec:restriction}
Recall that a policy profile is a collection of individual policies,
where each individual policy 
$\pi_i: O_i \mapsto A$ is a 
mapping from the agent's local states to available actions.
While computing this mapping,
it is sometime useful to partition the set of local states
into disjoint subsets and impose some restrictions on each subset based on the conditions that the mapping must satisfy.
Here, for each agent $i$, we consider partitioning its local states into two subsets:
$O_i = O_i^+ \cup O_i^-$, where $O_i^+ \cap O_i^- = \emptyset$.
We then compute a partial policy for each subset
$\pi_i^+: O_i^+ \mapsto A$ and $\pi_i^-: O_i^- \mapsto A$, and compose the full policy as $\pi_i = \pi_i^+ \cup \pi_i^-$.
Users can define their own ways of partition, e.g., let $O_i^-$ be the set of local states where agent~$i$ does not detect any other agents, and consequently, let $O_i^+$ consist of the remaining local states where agent $i$ can detect some others.
This will further allow users to specify human-understandable rules for either part (or both) to impose certain properties on the resulting computed policies.


\subsubsection{Action Preferences}
\label{sec:action_pref}
In this section, we investigate the feasibility of policies based on different choices
of $O_i^-$.
To customize a preference ordering over actions, we associate each action with an integral cost in given local state.
In principle, any distance measure can be used as a cost heuristic.
Here, we present empirical results based on the Manhattan distance from the agent to her goal, ignoring the subsequent actions of other agents.
Any action that moves the agent into obstacles or off the map will incur an infinite cost.
Note that several actions may have the same cost, i.e., the preference is not strict.
Therefore, each agent $i$ with such preferences intends to take actions in a self-interested manner under the local states within $O_i^-$.
We present three scenarios of choosing $O_i^-$, each of which accordingly defines a different heuristic.
\begin{enumerate}
    \item \textit{Default action} scenario. Let each $O_i^-$ be the set of local states where agent $i$ have not detected the others in its sensor.
    \item \textit{Last-minute coordination} scenario. Let each $O_i^-$ be the set of local states where agent $i$ are more than two blocks away from the others. Note that two agents that are no more than two blocks away may collide into each other if any subsequent coordination is made.
    \item \textit{Myopic preference} scenario. Let each $O_i^- = O_i$, i.e. the whole set of local states. This is the most aggressive heuristic that only allows an agent to pick one of the tied actions with the least cost. 
\end{enumerate}
Due the limited space, we postpone the rather intuitive encodings for these heuristics to Appendix~\ref{app:action_pref}.
One can notice that, from the first scenario to the third, larger portions of the partition are assigned to each $O_i^-$, and therefore, stronger constraints are imposed on the policies.

\begin{table}
\small
\centering
\begin{tabular}{lr}
\toprule
Scenario (map, \#agents)    & \#feasible/\#total \\
\midrule
\textit{\textbf{Sensor 1}} \\
$\#1 ~(6\times6, 2)$       & 8/1260 \\
\vspace{-3mm}
\\
\textit{\textbf{Sensor 2 or 3}} \\
$\#1 ~(6\times6, 2)$       & 1260/1260 \\
$\#2 ~(6\times6, 2)$       & 1260/1260  \\
$\#3 ~(5\times6, 2)$       & 192/870    \\
$\#3 ~(6\times6, 2)$       & 244/1260   \\
$\#3 ~(6\times7, 2)$       & 300/1722    \\
\bottomrule
\end{tabular}
\caption{Feasible policies may or may not exist under certain heuristics.
\#1, \#2, and \#3 correspond to
scenarios \textit{default action}, \textit{last-minute coordination},
and \textit{myopic preferences}, respectively.}
\label{tab:restriction}
\end{table}

First, we test whether feasible policies still exist under each preference heuristic, with some results reported in Table \ref{tab:restriction}.
For each tested configuration, we examine all possible goal profiles,
e.g., $36\times35 = 1260$ proper goal profiles for a $6\times6$ empty map with two agents.
Notice that, for the \textit{default action} and \textit{last-minute coordination} scenarios,
\textit{sensors 2} can already solve all configurations.  However, it may be too aggressive for the agents to resolve collisions under the \textit{myopic preference} heuristic, as indicated by several failed cases. One can also refer to some detailed elaboration in Appendix~\ref{apd:elabtab4}.

\begin{figure}[!ht]
    \centering
    \includegraphics[scale = 0.4]{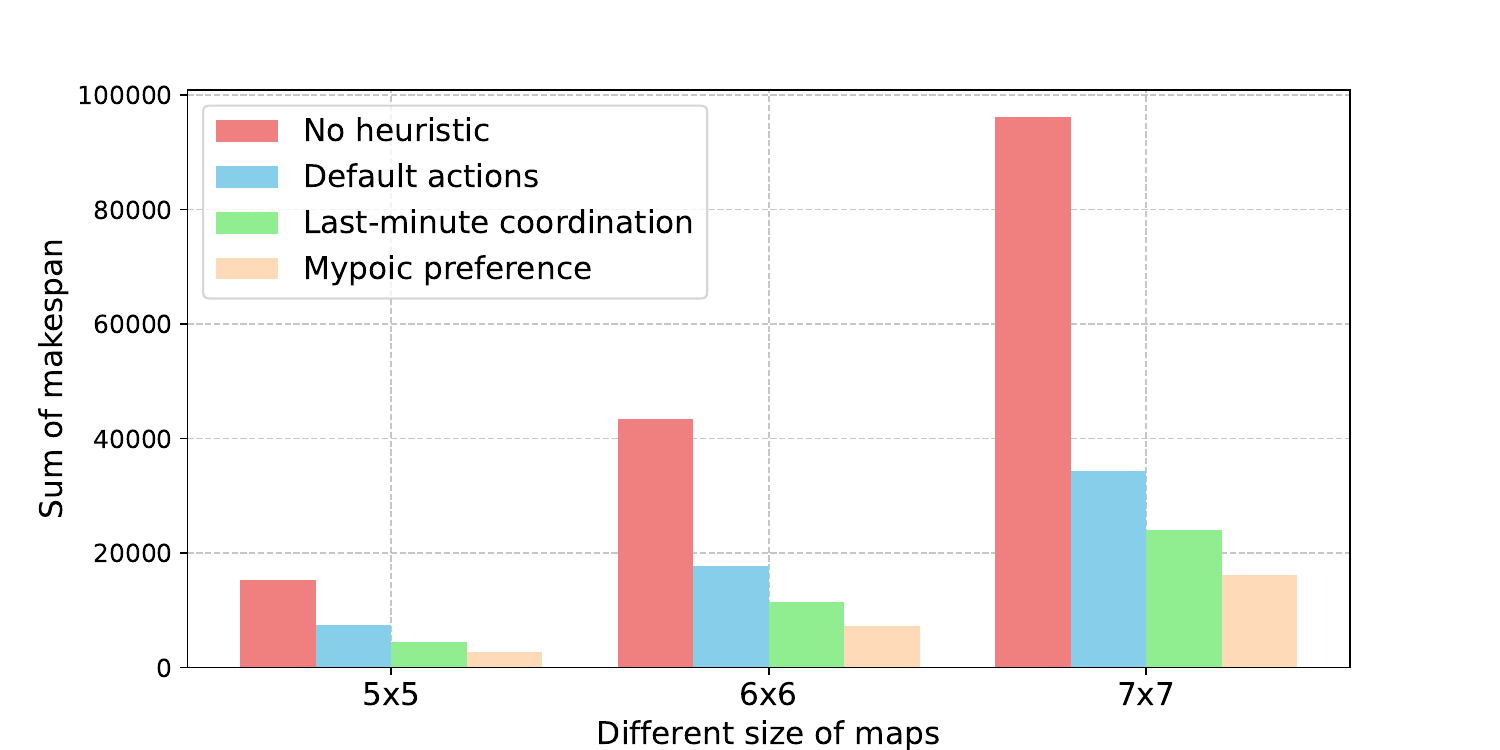}
    \caption{\textit{Sum-of-makespan} evaluated on different maps with different preference heuristics.
    }
    \label{fig:sat_makespan}
\end{figure}

Secondly, it is both natural and intriguing to investigate how much the quality of a feasible policy can be improved solely through the use of such action preferences, without any explicit optimization statement included in the program.
To this end, we evaluate the aforementioned metric \textit{sum-of-makespan} for the \textit{first} feasible policy found by the ASP solver across different maps for each of the scenarios mentioned above.
As shown in Figure \ref{fig:sat_makespan}, significant improvements in the computed feasible policies can be clearly observed, if one exists.
An insightful takeaway is that, for one,
users may want to design certain single-agent heuristics over $O_i^-$ to effectively prune the search space of feasible policies;
for another, it might be safer to make $O_i^-$ only a subset $O_i$, allowing some space for the agents to coordinate among themselves.

Lastly, as a supplement, we provide some elaboration on when and why no feasible policy exists, drawing inspiration from the experimental results in the \textit{myopic preference} scenario.

\begin{definition}[Crossroads and Streets]
In a 2D grid map, given a pair of goals $g_1 = (x_1, y_1)$ and $g_2 = (x_2, y_2)$, we call the locations $(x_1, y_2)$ and $(x_2, y_1)$ crossroads.
If the crossroads are not on the boundary of the given map layout, we call them interior crossroads. And also, $g_1$ and $g_2$ are said to be on the same street if $x_1 = x_2 \lor y_1 = y_2$.
\label{dfn:crossroadsandstreets}
\end{definition}

\begin{theorem}
	Given a map without obstacles, a feasible policy profile exists for two agents with sensors of range two, if their goals forms no interior crossroads and are not on the same street.
\label{thm::existence}
\end{theorem}

\begin{figure}[!ht]
\centering
\includegraphics[width=0.25\linewidth]{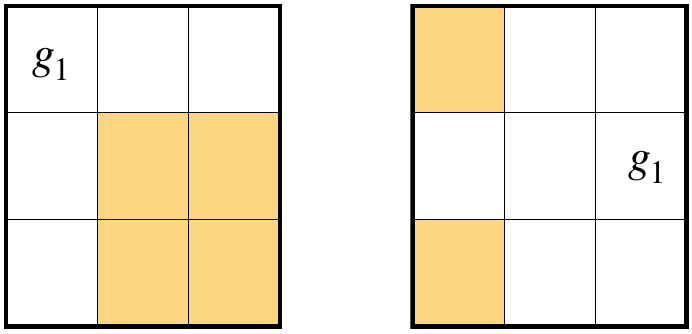}
\caption{Orange cells denote the choices of $g_2$ that lead to feasible policies.}
\label{fig:thm2eg}
\end{figure}

We note that the above theorem preliminarily characterizes a \textit{sufficient} condition for the existence of feasible policies in two-agent cases and provides an illustrative example in Figure~\ref{fig:thm2eg}.
The formal proof is deferred to Appendix~\ref{apd:theorem2}, while we provide an application as follows.
By applying the above theorem to a $6\times 6$ map with two agents, one can use simple combinatorics to calculate that there are $4\times 4\times 4 \times 2 + 5\times 6\times 2\times 2 - 4 = 244$ goal profiles that satisfy the condition, which matches the actual experimental results for the \textit{myopic preference} scenario in Table~\ref{tab:restriction} (similarly for 192 for $5\times 6$ configurations and 300 for $6\times 7$ configurations). As the \textit{myopic preference} heuristic imposes additional constraints on the original problem, then this feasibility result must hold in general without these constraints.

\subsubsection{Traffic Rules}
In addition to those single-agent heuristics developed on $O_i^-$,
we also intend to indentify some common behavioral patterns on $O_i^+$ , inspired by real-world traffic systems.
When a driver is approaching traffic lights, her driving decisions are made
according to certain traffic rules based on her relative position to the other cars.
In this context, a set of traffic rules serves as a common protocol that all drivers must obey
and is viewed from a relative frame.
In real-world situations, designers first come up with certain rules and
then make all drivers aware of them.
In this section, we investigate this idea in a reversed manner by asking the question:
\textit{
Ensuring the whole system to work, is there any underlying traffic rule that can be found?}
Let $f: \bigcup_i O_i^+ \mapsto O^+$ denote a function that transforms the local states where agents detect each other into those viewed from a unified relative frame.
A \textit{traffic rule} $tr: O^+ \mapsto A$
then maps these transformed local states to available actions,
and is adopted by all agents as common knowledge.
Once the complementary policy $\pi_i^-: O_i^- \mapsto A$ is determined,
the final policy profile can be collectively formed as
$\{\pi_i\}_{i\in N} = \{\pi_i^- \cup (tr \circ f) \}_{i\in N}$.
We explore two alternatives for encoding these traffic rules, with detailed encoding deferred to Appendix \ref{app:traf}.
\begin{enumerate}
	\item The traffic rule depends on both the location of the agent and its relative observations of others.
	One can opt to impose the \textit{default action} heuristic on those local states
where agents do not detect any others.
In this case, all agents are actually adopting the same way of routing (subject to different goals)
when they have not encountered each other, and then follow the same protocol when they do meet.
As shown in Table \ref{tab:traffic}, even when \textit{default actions} are preferred,
certain traffic rules can still be identified with the help of sensors with a range of only two.
	
	\item The traffic rule depends solely on the agent's relative observations of others, regardless of its own location.
	Surprisingly, completely opposite results are revealed in Table~\ref{tab:traffic}.
No such traffic rule is identified, even when we do not require agents to follow the \textit{default action} heuristic and equip them with the most powerful sensors (i.e., sensor 5, which can cover the entire $6\times 6$ map).
One reasonable explanation is that, unlike real-world traffic systems, our artificial map is bounded, and thus, agents are expected to behave differently as they approach the boundary, rather than behaving the same way as they do in the interior of the map.
\end{enumerate}

\begin{table}[!ht]
\small
\centering
\begin{tabular}{lrr}
\toprule
Scenario (map, agents)    & \#found/\#total  \\
\midrule
\textit{\textbf{Sensor 2, 3, 4 or 5}} \\
\#1 $(6\times6, 2)$       & 1260/1260 \\
\#1 $(6\times6, 2)$+default        & 1260/1260 \\
\#2 $(6\times6, 2)$       & 0/1260   \\
\#2 $(6\times6, 2)$+default       & 0/1260  \\
\bottomrule
\end{tabular}
\caption{Certain traffic rules exist when they depend on agents' locations, but no such traffic rules can be found if they are independent of agents' locations.
For both alternatives,
For both alternatives, we verify the existence of both the vanilla version and the \textit{default action} version.}
\label{tab:traffic}
\end{table}

\begin{table*}[!tp]
\small
\centering
\begin{tabular}{@{}l|rrr|rrr|rrr|l@{}}
\toprule
Map (\#agents)             & \multicolumn{3}{c|}{from scratch}                                                  & \multicolumn{3}{c|}{default action}                                                & \multicolumn{3}{c|}{last-minute coordination}                                      & \multicolumn{1}{l}{}           \\ \midrule
\multicolumn{1}{c|}{}      & \multicolumn{1}{c}{start} & \multicolumn{1}{c}{end} & \multicolumn{1}{c|}{$\downarrow$} & \multicolumn{1}{c}{start} & \multicolumn{1}{c}{end} & \multicolumn{1}{c|}{$\downarrow$} & \multicolumn{1}{c}{start} & \multicolumn{1}{c}{end} & \multicolumn{1}{c|}{$\downarrow$} & \multicolumn{1}{c}{$\downarrow^{total}$} \\ \midrule
\textit{\textbf{Sensor 2}} &           &          &            &           &           &             &               &              &                & \multicolumn{1}{l}{} \\
$3\times3$ (2)             & 360       & 181      & 49.72\%    & 415       & 263       & 36.63\%     & 264           & 203          & 23.11\%        & 43.61\%              \\
$4\times4$ (2)             & 4298      & 2732     & 36.44\%    & 2653      & 2473      & 6.78\%      & 1250          & 1241         & 0.72\%         & 71.13\%              \\
$5\times5$ (2)             & 15248     & 15149    & 0.65\%     & 7349      & 7302      & 0.64\%      & 4436          & 4422         & 0.32\%         & 71.00\%              \\
$6\times6$ (2)             & 43355     & 41007    & 5.42\%     & 17734     & 17331     & 2.27\%      & 11420         & 11420        & 0.00\%         & 73.66\%              \\
                           &           &          &            &           &           &             &               &              &                & \multicolumn{1}{l}{} \vspace{-3mm}\\
\textit{\textbf{Sensor 3}} &           &          &            &           &           &             &               &              &                & \multicolumn{1}{l}{} \\
$4\times4$ (2)             & 3391      & 3379     & 0.35\%     & 3368      & 3256      & 3.33\%      & 1462          & 1441         & 1.44\%         & 57.51\%              \\
$5\times5$ (2)             & 14962     & 14746    & 1.44\%     & 10423     & 10105     & 3.05\%      & 4605          & 4539         & 11.43\%        & 69.66\%              \\
$6\times6$ (2)             & 52530     & 45953    & 12.52\%    & 26873     & 26873     & 0.00\%      & 11544         & 11526        & 0.16\%         & 78.06\%              \\
                           &           &          &            &           &           &             &               &              &                & \multicolumn{1}{l}{} \vspace{-3mm}\\
\textit{\textbf{Sensor 4}} &           &          &            &           &           &             &               &              &                & \multicolumn{1}{l}{} \\
$5\times5$ (2)             & 22801     & 21950    & 3.73\%     & 14359     & 11277     & 21.46\%     & 4894          & 4882         & 0.25\%         & 78.59\%              \\
$6\times6$ (2)             & 54669     & 48316    & 11.62\%    & 29242     & 29238     & 0.01\%      & 11669         & 11628        & 0.35\%         & 78.73\%              \\ \bottomrule
\end{tabular}
\caption{Optimization results starting from different action heuristics.
The subtitle ``start'' means the sum-of-makespan of first feasible policy
while ``end'' means that of the last one, after which no better one has been found for 2 hours.
``Default action'' and ``last-minute coordination'' correspond to the respective heuristics in Section \ref{subsec:restriction}.}
\label{tab:opt_start_end}
\end{table*}

\subsection{Anytime Optimization}

When it comes to policy optimization,
the system designer can also specify human-understandable rules for $\pi_i^-$ in the first place,
and the optimization procedures are therefore only responsible for (optimally) ``completing'' the rest $\pi_i^+$.
Recall that we have designed several action heuristics for $\pi_i^-$
in Section~\ref{subsec:restriction}.
As reported in Table~\ref{tab:opt_start_end},
we experiment on optimizing the policy from scratch as well as with different action heuristics .
The program terminates after a sufficiently long period without finding a better solution.
This indicates that the improvement achieved through optimization is significantly less than that obtained by incorporating effective heuristics.
A lesson taken from this table is that, to compute better policies,
attention should be paid to identifying helpful constraints or properties compatible with the (near-)optimal policies,
rather than solely relying on the solver for the entire optimization process.



\section{Conclusion}
This paper presents the implementation of an ASP-based system for computing universal plans for any arbitrary 2D map and a group of partially observable agents with specific goals.
\textit{We emphasize again that our solution concept, which consists of a collection of decentralized policies, can suggest immediate actions whenever agents encounter any contingency. This is the key contribution that distinguishes our work from the existing literature, particularly in the area of MAPF.}
The system is used for comprehensive studies on the behavioral patterns of the agents.
With the help of ASP, user-specified preferences can be easily represented, thus improving the policy.
Furthermore, we highlight several potential directions and applications for our system.
\begin{itemize}
    \item Despite being able to solve larger problem instances compared to existing Dec-POMDP solvers,
	real-world applications may require even larger scales that are currently intractable for our system.
    One possible approach is to leverage our system to compute policies for (local) multi-agent coordination in sub-regions,
    and utilize other single-agent planners to guide agents while they are operating alone.

    \item Another valuable application is to integrate our rule-based system into a learning framework.
    Some work has explored a similar idea by manually designing \textit{acceleration rules} and \textit{safety rules} for single-agent RL tasks~\cite{gao2020embedding}.
    Additionally, our system provides insights into how to automate the generation of rules to support learning in multi-agent domains.
\end{itemize}

\section*{Acknowledgments}
We thank Jiuzhi Yu for his invaluable efforts and unwavering support in the initial stages of this work.
We also thank Yuxin Pan and the anonymous reviewers for their insightful comments.

\bibliographystyle{eptcs}
\bibliography{generic}

\clearpage
\appendix

\section{ASP Encoding for the Problem}
\label{apd:prob_encoding}

In this section, we show how to encode the problem by ASP.
For the simplicity of presentation, we present \texttt{p:- q}
in the meaning of $p \leftarrow q$. First, we encode the following facts:
\begin{itemize}
    \item \texttt{cell(X,Y)} and \texttt{block(X,Y)}
    encode the empty cells and obstacles on the map, respectively.
    \item \texttt{goali((GX,GY))}
    specifies the goal designated to agent $i$.
\end{itemize}
Actions are encoded as follows
    (similarly for \texttt{down}, \texttt{right}, \texttt{left} and \texttt{nil}).
\begin{lstlisting}[style=gringo]
action(X,Y,up):- cell(X,Y),  not  block(X,Y),
    cell(X-1,Y),  not  block(X-1,Y).
\end{lstlisting}
Then we encode the global state representation,
which is a snapshot of the whole system including the current positions and the respective
goals of all agents.
Note that all agents are supposed to be placed at an empty cell and not to overlap with each other.
\begin{lstlisting}[style=gringo]
gState(S):- S=(L1,...,Ln,G1,...,Gn), goal1(G1), ..., goaln(Gn),
    Li=(Xi,Yi), cell(Xi,Yi),  not  block(Xi,Yi), % for all i
    Li!=Lj. % for all distinct i,j -> no vertex conflict
\end{lstlisting}
Upon the definition of global states,
we then define the transition.
A global transition is triggered upon a collection of all agents' reported actions.
Note that any pair of agents cannot traverse the same passage, i.e. no immediate swap is allowed.
\begin{lstlisting}[style=gringo]
trans(S1,A1,...,An,S2):- gState(S1), gState(S2),
    S1=(L11,...,L1n,G11,...,G1n), S2=(L21,...,L2n,G21,...,G2n),
    L11=(X11,Y11), ..., L1n=(X1n,Y1n),
    L21=(X21,Y21), ..., L2n=(X2n,Y2n),
    move(X1i,Y1i,Ai,X2i,Y2i), % for all i
    (L1i,L1j)!=(L2j,L2i). % for all distinct i,j -> no edge conflict
\end{lstlisting}
To make the system goal-directed, we also need to define the global goal state.
\begin{lstlisting}[style=gringo]
goal_gState(S):- gState(S), S=(G1,...,Gn,G1,...,Gn),
    goal1(G1), ..., goaln(Gn).
\end{lstlisting}
As we mentioned earlier, each agent is only capable of observing others within her local FoV.
\begin{lstlisting}[style=gringo]
aState(AS):- AS=(Self,O2,...,On,Goal),
    Self=(X,Y), cell(X,Y),  not  block(X,Y),
    Goal=(Xg,Yg), cell(Xg,Yg),  not  block(Xg,Yg),
    near(Self,Oi). % for all i=2..n
\end{lstlisting}
The \texttt{near/2} predicate denotes the functionality of the sensor.
If the opponent is beyond the detection of the agent,
we let the constant \texttt{empty} be the placeholder for such a null position.
The constant \texttt{R} denotes the sensor range.
One can alternatively adopt any other distance measure, e.g. Manhattan or Euclidean distance.
For simplicity, our example adopts the $L_\infty$ metric as the distance measure, i.e. $dist(p, q) = \max\{|p_x - q_x|, |p_y - q_y|\}$.
\begin{lstlisting}[style=gringo]
near(Self,empty):- Self=(X1,Y1), cell(X1,Y1).
near(Self,Other):- Self!=Other, Self=(X1,Y1), Other=(X2,Y2),
    cell(X1,Y1),  not  block(X1,Y1), cell(X2,Y2),  not  block(X2,Y2),
    |X1-X2| <= R, |Y1-Y2| <= R.
\end{lstlisting}
As required,
each agent will not be incentivized to do any further move once she reaches the goal.
Thus, each agent should be capable of recognizing her own dedicated goal.
\begin{lstlisting}[style=gringo]
% for all i
goali_aState(AS):- aState(AS), AS=(Gi,_,...,_,Gi), goali(Gi). 
\end{lstlisting}
To map from global states to each agent's local states,
we then define the observation model.
Note that by \texttt{(L1,...,Ln)$\backslash$Self} we mean
to exclude the agent herself from the vector of all agents' locations. 
By \texttt{empty/Oi} we mean that the $i$-th one of the other agents
can be outside or inside the FoV.
Correspondingly, if \texttt{Oi} is not detected,
then we include \texttt{not near(Self,Oi)}, otherwise no leading \texttt{not}.
The observation model is in fact the most expensive one to encode in our system,
since we need to enumerate all possible $2^{n-1}$ situations of sensor detection for each agent,
thus, $n\cdot 2^{n-1}$ axioms in total.

\begin{lstlisting}[style=gringo]
obsi(S,AS):- gState(S), S=(L1,...,Ln,G1,...,Gn),
    Self=Li, (O2,...,On)=(L1,...,Ln)\Self,
    aState(AS), AS=(Self,empty/O2,...,empty/On,Gi),
    { not } near(Self,Oi). % for all i=2..n
\end{lstlisting}
On top of the above, we formulate each individual policy under two restrictions:
1) once an agent reaches her dedicated goal, she has to stop;
2) pick one single available action otherwise.
Our desired solution is consequently a deterministic one and
executable from the local perspective of an agent.
\begin{lstlisting}[style=gringo]
doi(AS,nil):- goali_aState(AS).
{doi(AS,A): avai_action(AS,A)}=1 :- aState(AS),
    AS=(Self,_,...,_,Goal), 
    Self != Goal, goali(Goal).
\end{lstlisting}
To ensure that the policy works for every contingency,
we inductively define the reachability of each global state given the policy.
Basically, a goal global state is trivially reachable to itself.
Inductively, any global state is reachable to the goal state
if it can, according to the policy, transit to a successor state that is reachable to goal state. 
\begin{lstlisting}[style=gringo]
reached(S):- goal_gState(S).
reached(S1):- gState(S1), trans(S1,A1,...,An,S2), reached(S2),
    obsi(Si,ASi), doi(ASi,Ai). % for all i
\end{lstlisting}
Finally, we include the integrity constraint to rule out infeasible policies.
\begin{lstlisting}[style=gringo]
:- gState(S),  not  reached(S).
\end{lstlisting}

\section{Existence of Policies for Some Manually Designed Maps}
\label{apd:manual}

\begin{figure}[h]
    \centering
    \includegraphics[width=15mm]{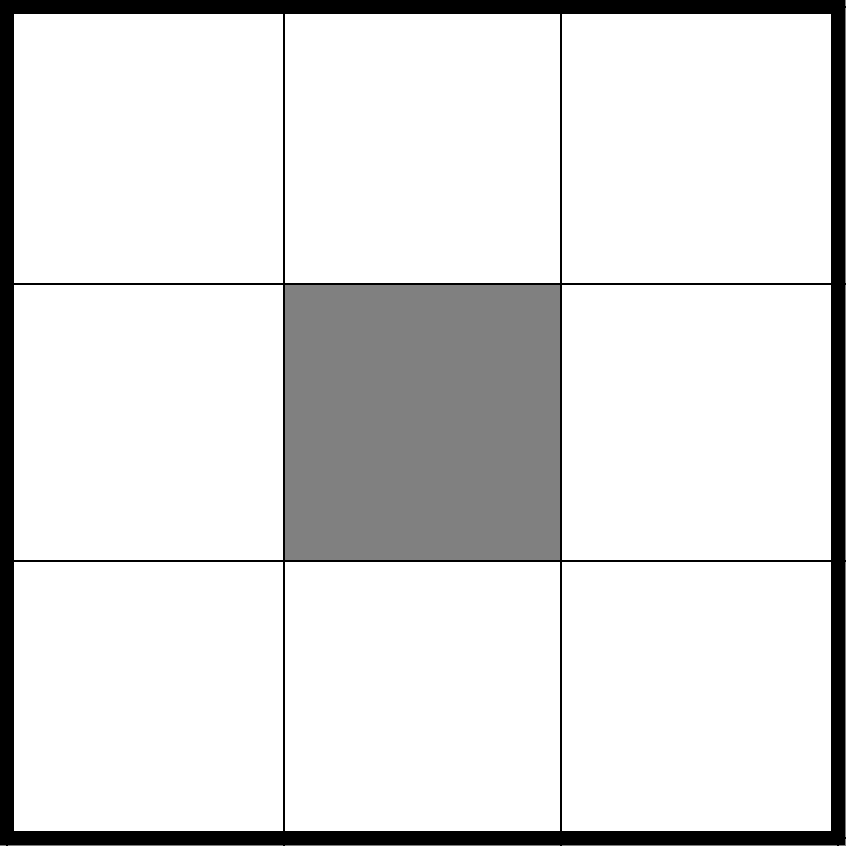}
    \includegraphics[width=25mm]{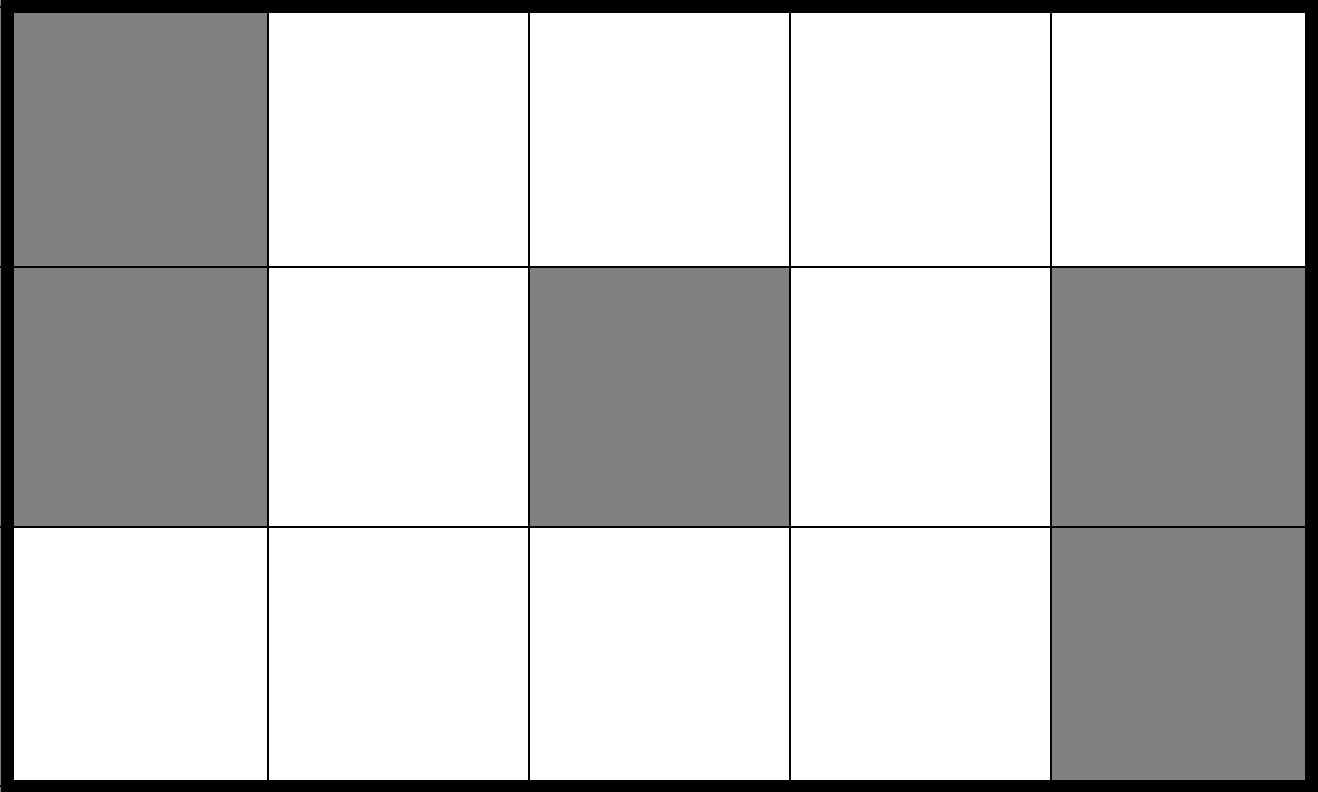}
    \includegraphics[width=25mm]{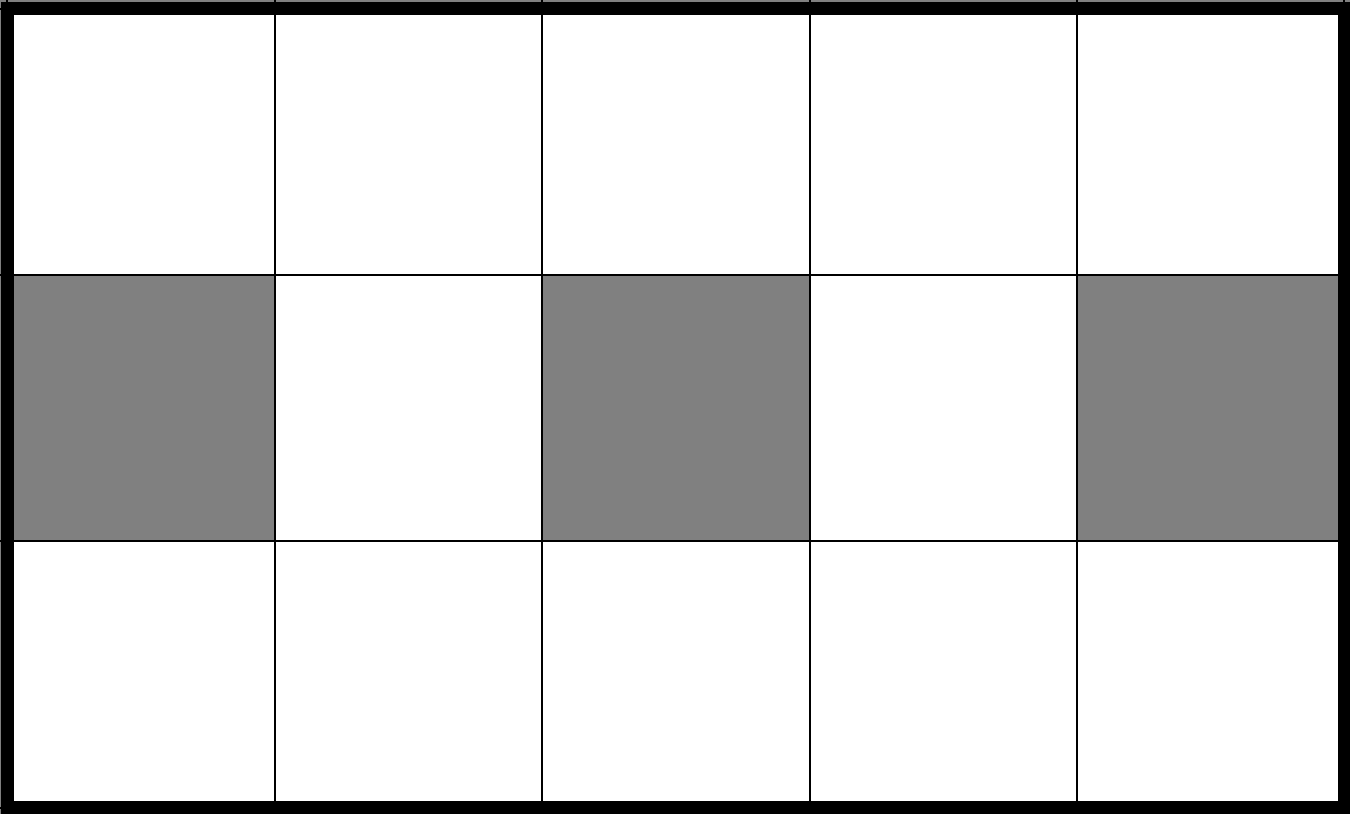}\\
    \includegraphics[width=15mm]{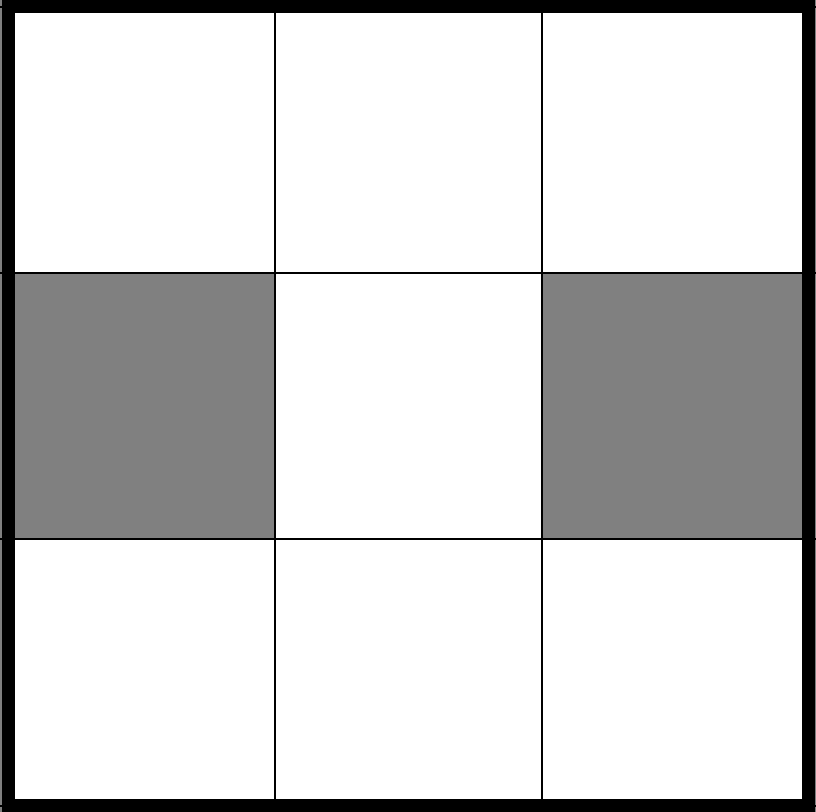}
    \includegraphics[width=15mm]{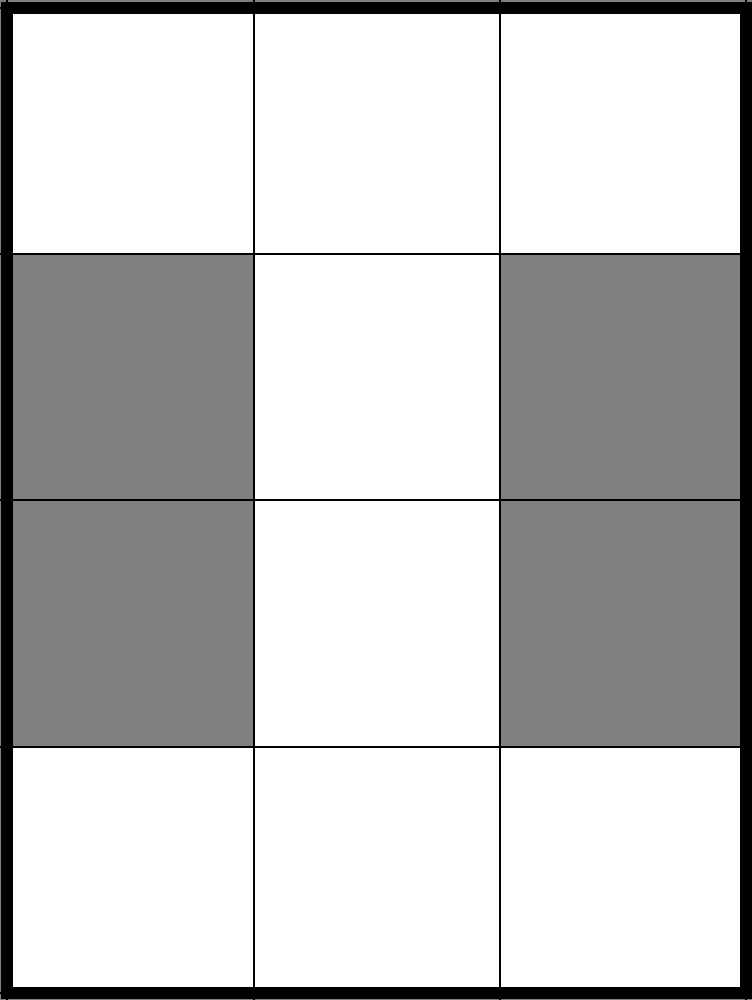}
    \includegraphics[width=35mm]{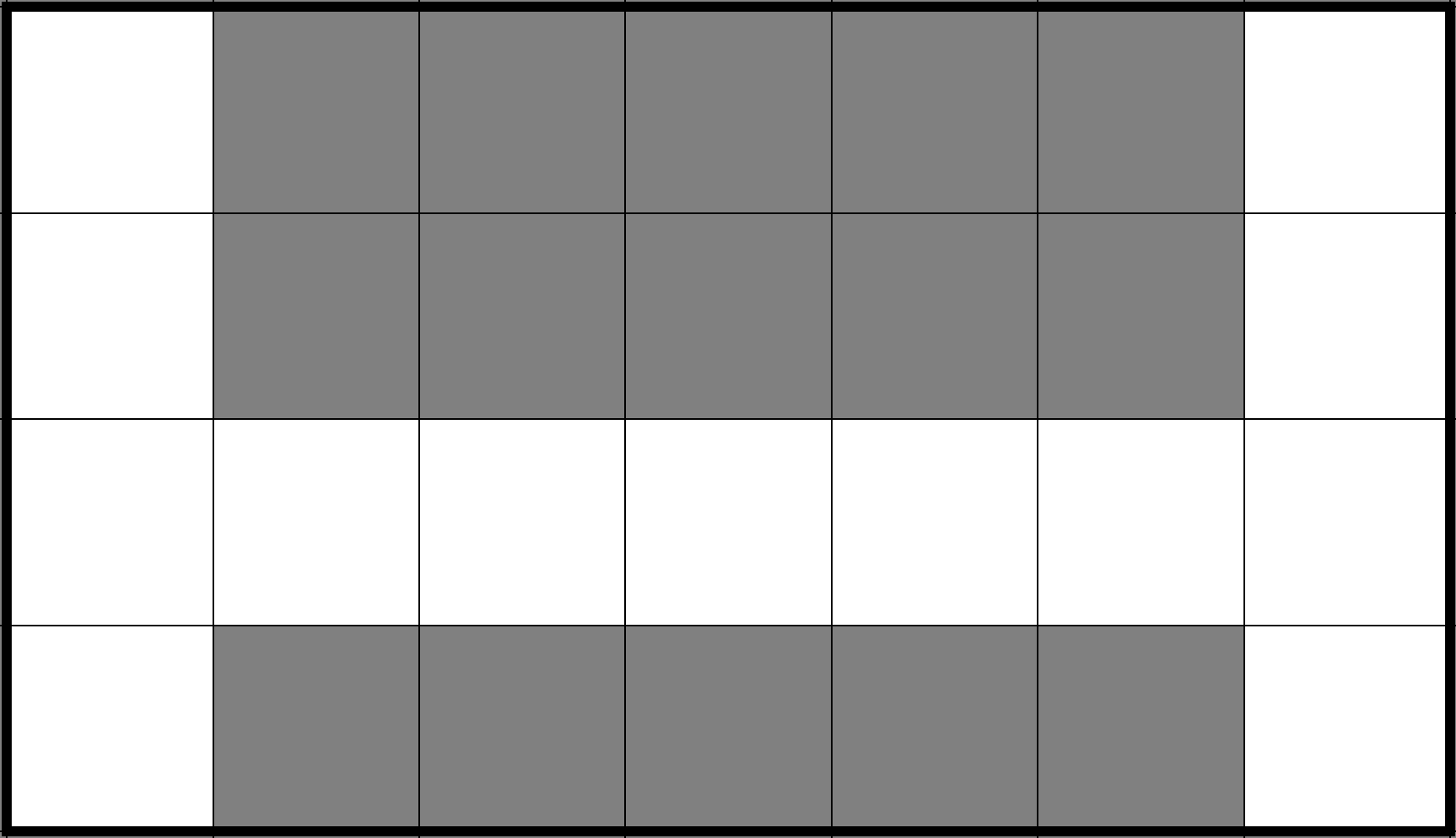}
    \caption{Manually designed layouts for the results reported in Table \ref{tab:sensor_man}.
	From the left to the right and the top to the bottom,
	layouts are identified as \texttt{id\_1} to \texttt{id\_6}.}
    \label{fig:map_id}
\end{figure}

\begin{table}[ht]
\centering
\begin{tabular}{lrrr}
\toprule
Map (\#agents)    & sensor 1  & sensor 2  & full sensor    \\
 & \#feasible/\#proper/\#total & \#feasible/\#proper/\#total & \#feasible/\#proper/\#total \\
\midrule
$\texttt{id\_1}(2)$  & 28/56/56  & 56/56/56     & no need \\
$\texttt{id\_2}(2)$  & 46/56/90  & 56/56/90     & 56/56/90 \\
$\texttt{id\_3}(3)$  & 0/168/1320  & 168/168/1320   & 168/168/1320 \\
$\texttt{id\_4}(3)$  & 0/24/210   & 24/24/210     & no need \\
$\texttt{id\_5}(3)$  & 0/24/336   & 24/24/336     & 24/24/336 \\
$\texttt{id\_6}(3)$  & 0/24/1716   & 24/24/1716     & 24/24/1716 \\
\bottomrule
\end{tabular}
\caption{Feasibilities with different sensors, for these manually designed layouts in Figure \ref{fig:map_id}.}
\label{tab:sensor_man}
\end{table}

To investigate
how the existence of feasible policies may depend on the capability of sensors,
we conduct experiments on six manually designed map layouts,
as shown in Table \ref{tab:sensor_man}.
We evaluate the feasibility over all possible \textit{proper} goal profiles
for agents with sensors of different ranges.
Detailed map layouts are shown in Figure \ref{fig:map_id}.
Map \texttt{id\_1} and \texttt{id\_2} are designed with roundabouts for two-agent cases,
and map \texttt{id\_3} -- \texttt{id\_6} are designed to further include channels for three agent cases.
Notice that except for map \texttt{id\_1} and \texttt{id\_4},
agents need sensors of larger range to cover the whole map
but actually sensors of range two already suffice.
\textit{Full sensors} are the ones that can cover the whole map.

\section{ASP Encoding for Action Preferences}
\label{app:action_pref}

\subsection{The \texttt{cost} Predicate}
\label{sec::axioms}
Let the predicate \verb-cost(AS,A,C)- define the Manhattan distance \verb-C-
from the local state \verb-AS- to the goal by taking action \verb-A-.
The Manhattan distance is computed from the perspective of the agent herself,
without any prescription of the actions that will be taken by other agents.

 Below shows an example how we encode Manhattan distance for two agents
 (encodings for $n>2$ agents can be easily generalized).
 If the action will not lead to any immediate collision,
 the cost will simply be the Manhattan distance towards the goal.
 \begin{lstlisting}[style=gringo]
cost(AS,Action,C):-
    aState(AS), AS=(Self,empty,Goal),
    Self=(X,Y), Goal=(Xg,Yg),
    move(X,Y,Action,Xs,Ys),
    C=1+|Xg-Xs|+|Yg-Ys|.

cost(AS,Action,C):-
    aState(AS), AS=(Self,Other,Goal),
    near(Self,Other),
     not  nextTo(Self,Action,Other),
    Self=(X,Y), Goal=(Xg,Yg),
    move(X,Y,Action,Xs,Ys),
    C=1+|Xg-Xs|+|Yg-Ys|.

cost(AS,Action,INFTY):-
    aState(AS), AS=(Self,Other,Goal),
    near(Self,Other),
    nextTo(Self,Action,Other),
    Self=(X,Y), Others2=(Xs,Ys),
    move(X,Y,Action,Xs,Ys).

nextTo(Loc1, Action, Loc2) :- Loc1 = (X1, Y1), Loc2 = (X2, Y2),
    cell(X1, Y1), not block(X1, Y1),
    cell(X2, Y2), not block(X2, Y2),
    move(X1, Y1, Action, X2, Y2).
 \end{lstlisting}
 Otherwise, agents may run into walls or go beyond the map, also incurring a cost of infinity.
 \begin{lstlisting}[style=gringo]
 cost(AS,Action,INFTY):-
     aState(AS), isAction(Action),
      not  avai_action(AS,Action).
 \end{lstlisting}
 Note that this cost metric is defined in such a fashion 
 that an agent prefers actions with the least Manhattan distance
 without regard to the other agents' true subsequent actions.

\subsection{Three Types of Preferences}

For \underline{\textit{default actions}},
we modify the policy
definition\footnote{without particular specification, we usually keep the stop-at-goal rule unchanged.} as
\begin{lstlisting}[style=gringo]
{doi(AS,A):avai_action(AS,A)}=1 :- 
    aState(AS), AS=(Self,Other,Goal),
    near(Self,Other), Other!=empty,
    Self!=Goal, goali(Goal).

{doi(AS,A): avai_action(AS,A),
    cost(AS,A,C0),
    cost(AS,up,C1), ..., cost(AS,nil,C5),
    C0<=C1,...,C0<=C5}=1 :- 
    aState(AS), AS=(Self,empty,Goal),
    Self!=Goal, goali(Goal).
\end{lstlisting}

\noindent
For \underline{\textit{last-minute coordination}},
we further sophisticate the policy definition as
\begin{lstlisting}[style=gringo]
{doi(AS,A):avai_action(AS,A)}=1 :- 
    aState(AS), AS=(Self,Other,Goal), 
    near(Self,Other), Other!=empty,
    lastmin(Self,Other),
    Self!=Goal, goal2(Goal).

{doi(AS,A):avai_action(AS,A),
    cost(AS,A,C0),
    cost(AS,up,C1), ..., cost(AS,nil,C5),
    C0<=C1, ..., C0<=C5}=1 :- 
    aState(AS), AS=(Self,Other,Goal),
    near(Self,Other), Other!=empty,
     not  lastmin(Self,Other),
    Self!=Goal, goali(Goal).

{doi(AS,A):avai_action(AS,A),
    cost(AS,A,C0),
    cost(AS,up,C1), ..., cost(AS,nil,C5),
    C0<=C1, ..., C0<=C5}=1 :- 
    aState(AS), AS=(Self,empty,Goal),
    Self!=Goal, goali(Goal).
\end{lstlisting}
where the \texttt{lastmin} predicate defines the moments
when two agents are about to collide with each other if both make inappropriate actions, i.e., they are two units of Manhattan distance away from each other.
\begin{lstlisting}[style=gringo]
lastmin(P1,P2):-
    P1=(X1,Y1), P2=(X2,Y2),
    cell(X1,Y1), cell(X2,Y2),
    |X1-X2|+|Y1-Y2|<=2.
\end{lstlisting}

\noindent
For \underline{\textit{myopic preferences}},
we do not have to discuss that many sub-cases, therefore, can simply write
\begin{lstlisting}[style=gringo]
{doi(AS,A):avai_action(AS,A),
    cost(AS,A,C0),
    cost(AS,up,C1), ..., cost(AS,nil,C5),
    C0<=C1, ..., C0<=C5}=1 :- 
    aState(AS), AS=(Self,_,Goal),
    Self!=Goal, goali(Goal).
\end{lstlisting}

\section{ASP Encoding for Traffic Rules}
\label{app:traf}
\underline{\textit{For the first alternative}}, the policy definitions for $\pi_i^+$ are modified as follows. Note that we omit the definitions for $\pi_i^-$ here. One can opt to impose the \textit{default action} heuristic on those local states where agents have not detected any of the others.
\begin{lstlisting}[style=gringo]
{traffic(Self,Rel_pos,A): avai_action(AS,A)}=1 :- 
    aState(AS), AS=(Self,Other,Goal),
    near(Self,Other), Other!=empty,
    Self=(X1,Y1), Other=(X2,Y2),
    Rel_pos=(X2-X1,Y2-Y1),
    Self != Goal.

{doi(AS,A): traffic(Self,Rel_pos,A)}=1 :- 
    aState(AS), AS=(Self,Other,Goal),
    near(Self,Other), Other!=empty,
    Self=(X1,Y1), Other=(X2,Y2),
    Rel_pos=(X2-X1,Y2-Y1),
    Self!=Goal, goali(Goal).
\end{lstlisting}

\noindent
\underline{\textit{For the second alternative}}, we instead include the following rules. And also, one can implement $\pi_i^-$ as the \textit{default action} version.
\begin{lstlisting}[style=gringo]
{traffic(Rel_pos,A): avai_action(AS,A)}=1 :- 
    aState(AS), AS=(Self,Other,Goal),
    near(Self,Other), Other!=empty,
    Self=(X1,Y1), Other=(X2,Y2),
    Rel_pos=(X2-X1,Y2-Y1),
    Self!=Goal.

{doi(AS,A): traffic(Rel_pos,A)}=1 :- 
    aState(AS), AS=(Self,Other,Goal),
    near(Self,Other), Other!=empty,
    Self=(X1,Y1), Other=(X2,Y2),
    Rel_pos=(X2-X1,Y2-Y1),
    Self!=Goal, goali(Goal).
\end{lstlisting}

\section{Elaboration on Table~\ref{tab:restriction}}
\label{apd:elabtab4}
\begin{figure}[!ht]
\centering
\includegraphics[width=0.5\linewidth]{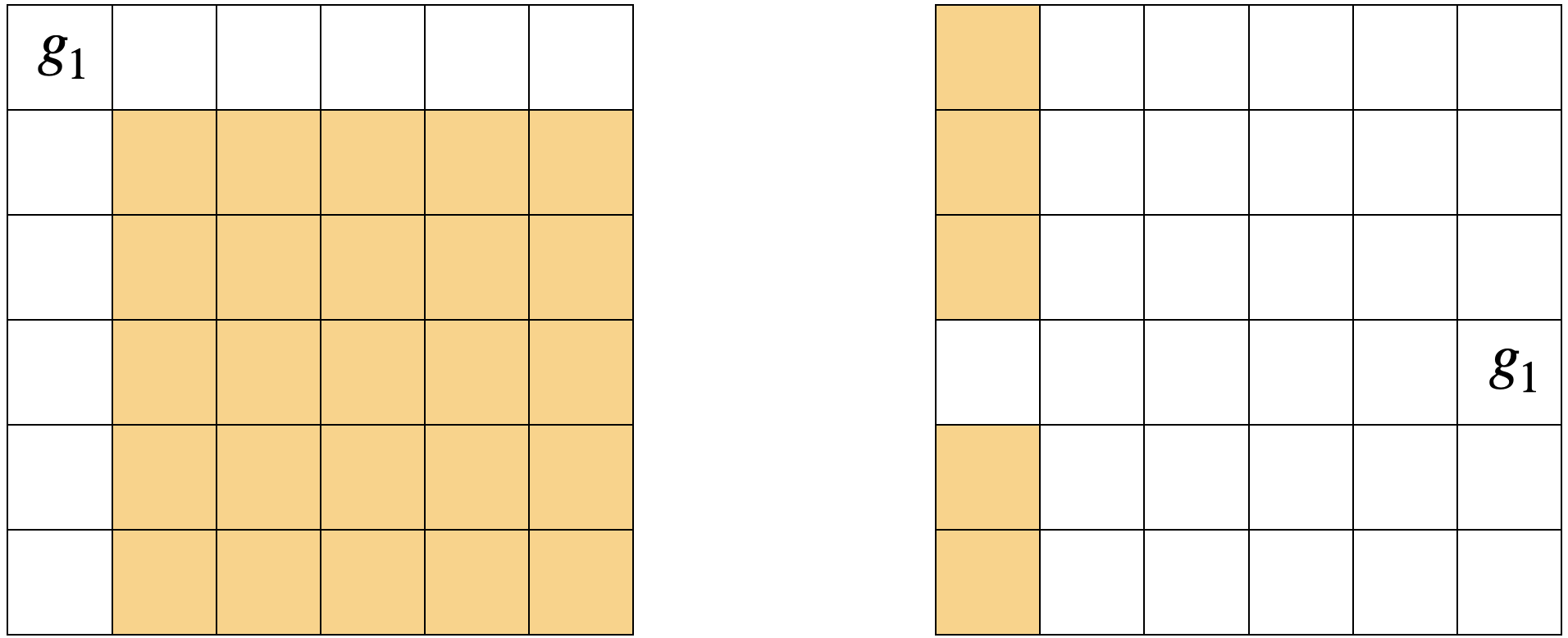}
\caption{Orange cells denote possible $g_2$'s that lead to feasible policies. (a) Left: the corner case; (b) Right: the boundary case (other than corners).}
\label{fig::exp}
\end{figure}

One can easily run experiments to check if there always exist any feasible universal policies in small domains, e.g. two agents in 6-by-6 grid worlds with no obstacles.
Our experiments on 6-by-6, 5-by-6 and 6-by-7 empty grid maps with two agents have shown that feasible policies exist for goal profiles that are in one of the following conditions:
\begin{enumerate}
	\item If one of the goal is at the corner, the other goal can be at any cell within the submap induced by eliminating the column and row where the former goal is located. As shown in Figure \ref{fig::exp}(a), suppose $g_1$ is at the upper left corner, then $g_2$ can be any one of cells in the orange space.
	\item If one of the goal is at one of the four boundaries other than the corner, the other goal can only be at any cell that are located at the opposite boundary other than the cell that are in the same line. As shown in Figure \ref{fig::exp}(b), suppose $g_1$ is at the rightmost boundary, then $g_2$ can be any one of cells in the orange space. Same situations happen when $g_1$ is at the one of other three boundaries.
\end{enumerate}
Take the 6-by-6 map with two agents as an example, we have totally $(6\times 6) \times (6\times 6 - 1) = 1260$ proper goal profiles.
According to the description above, we have $4\times 4\times 4 \times 2 + 5\times 6\times 2\times 2 - 4 = 244$ goal profiles that allow feasible policies to exist, which matches the actual experimental results (192 for 5-by-6 layouts, 244 for 6-by-6 layouts, and 300 for 6-by-7 layouts).

\section{Elaboration on Theorem~\ref{thm::existence}}
\label{apd:theorem2}

Further, based on experimental findings, we will then formally characterize the results for maps of any size with two agents, starting with a few definitions.

\begin{figure}[!ht]
\centering
\includegraphics[width=0.6\linewidth]{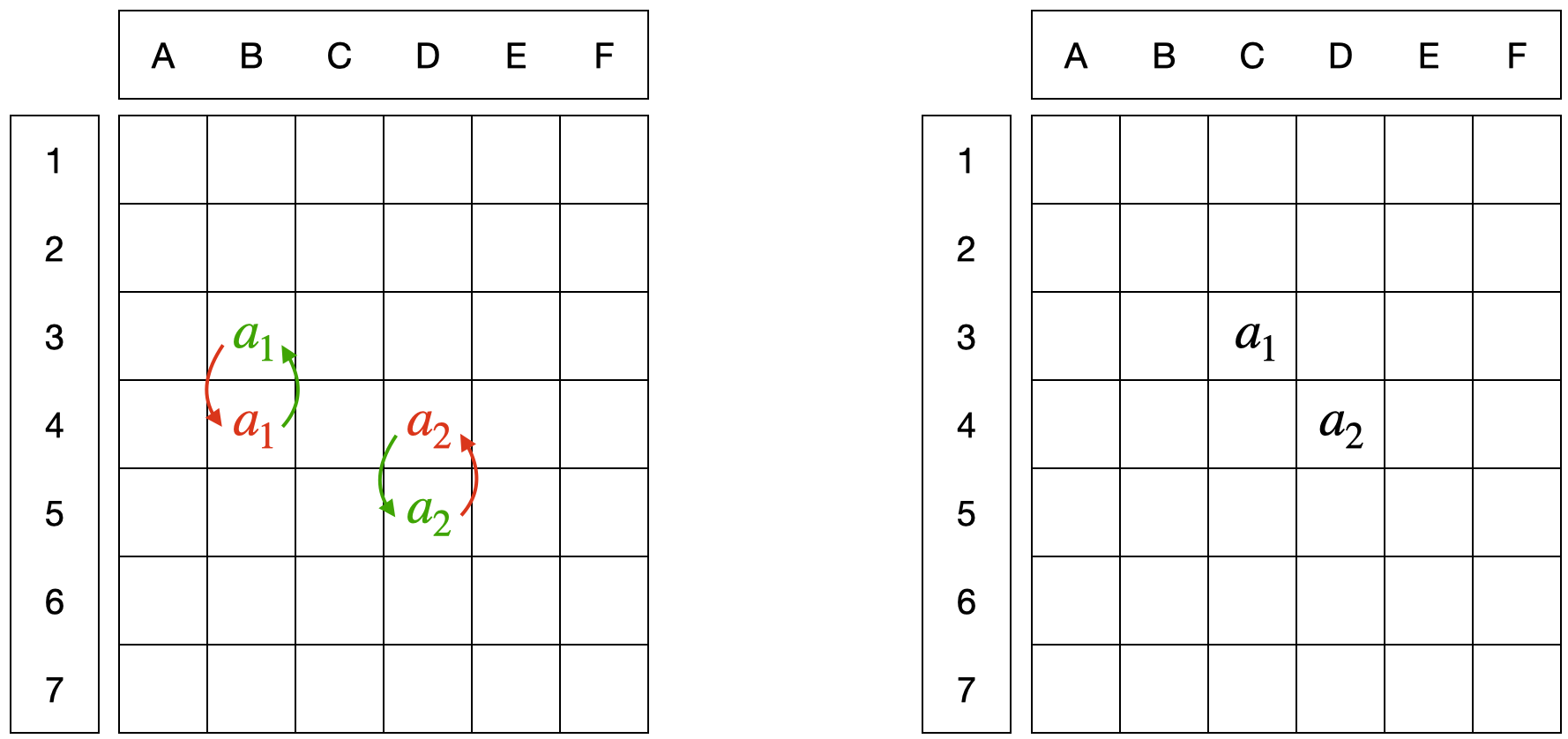}
\caption{(a) Left: an example of a live lock; (b) Right: an example of a dead lock if both stop.}
\label{fig::locks}
\end{figure}

\begin{definition}
	(Locks). If a group of agents repeatedly go through the same set of global states, they are said to be in locks. This particular set of global states are called a lock set. If the lock set contains more than one states, agents are said to be in a live lock. Otherwise, they are said to be in a dead lock. 
\end{definition}
For example, Figure \ref{fig::locks}(a) has shown a case of a live lock. Agents start from red ones, simultaneously go to green ones and get back to red ones. The lock set contains two states $\{[(4, B), (4, D)], [(3, B), (5, D)]\}$. An example of deadlock is also shown in Figure \ref{fig::locks}(b) if two agents simultaneously decide to stop, then they will consequently both stop forever since their policies only depend the current local states, leading to a lock set of $\{[(3, C), (4, D)]\}$.

\begin{proposition}
	The rule that encodes myopic preferences will rule out any live lock.
\label{prop::livelock}
\end{proposition}
The proposition is way clear to prove. For a live lock to exist, there should be at least two global states in the lock set, which means at least one of the agents would visit her previous local state (different from the current one). Nevertheless, according to the definition of \textit{myopic preferences}, even if an agent's most preferred action is blocked by another agent, she would stop rather than going backwards, which results in a less Manhattan distance.

\begin{figure}[!ht]
\centering
\includegraphics[width=0.6\linewidth]{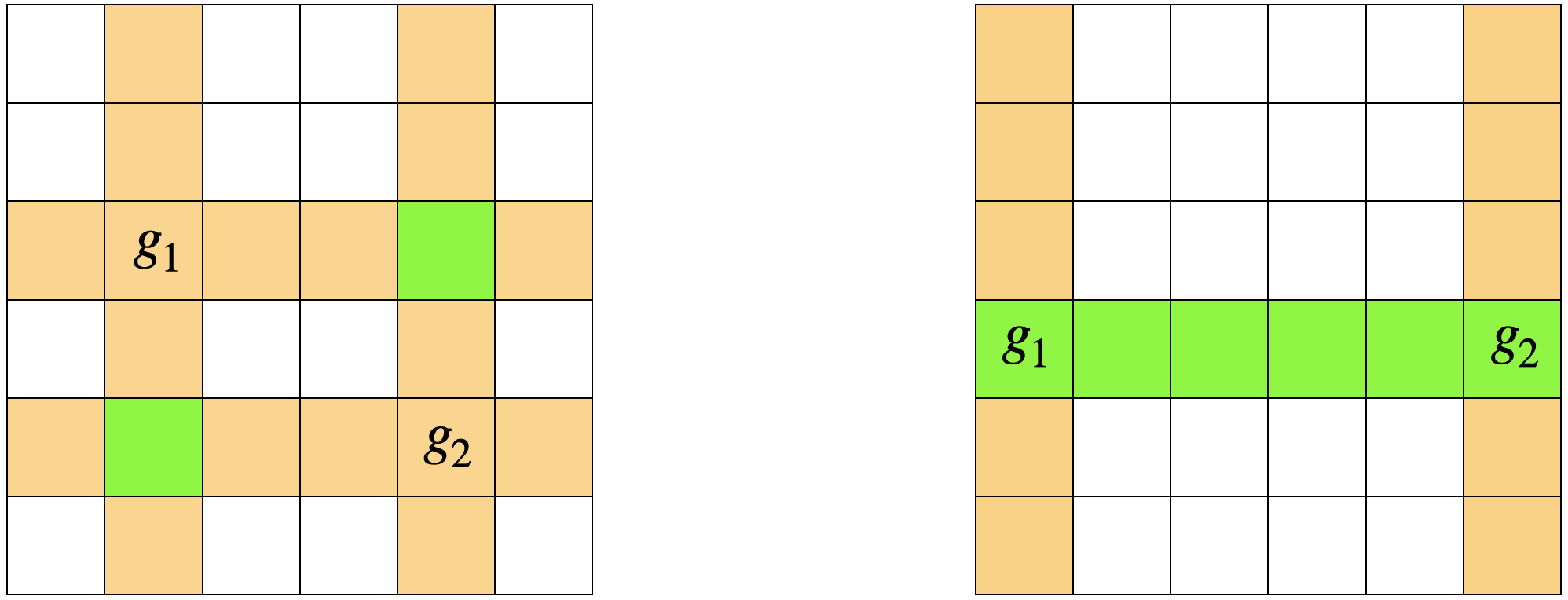}
\caption{Green cells denote crossroads. (a) Left: a typical case of crossroads; (b) Right: a degraded case when two goals are in the same column/row.}
\label{fig::crossroads}
\end{figure}

Recall that in Definition~\ref{dfn:crossroadsandstreets}, we define \textit{crossroads} and \textit{streets}.
To illustrate with an example, Figure~\ref{fig::crossroads}(a) has shown a typical case where cells in orange are the columns/rows where goals are located, and the orange cells are the consequent interior crossroads.
In Figure~\ref{fig::crossroads}(b), the two goals are in the same street marked in green as well.

\begin{figure}[!ht]
\centering
\includegraphics[width=1\linewidth]{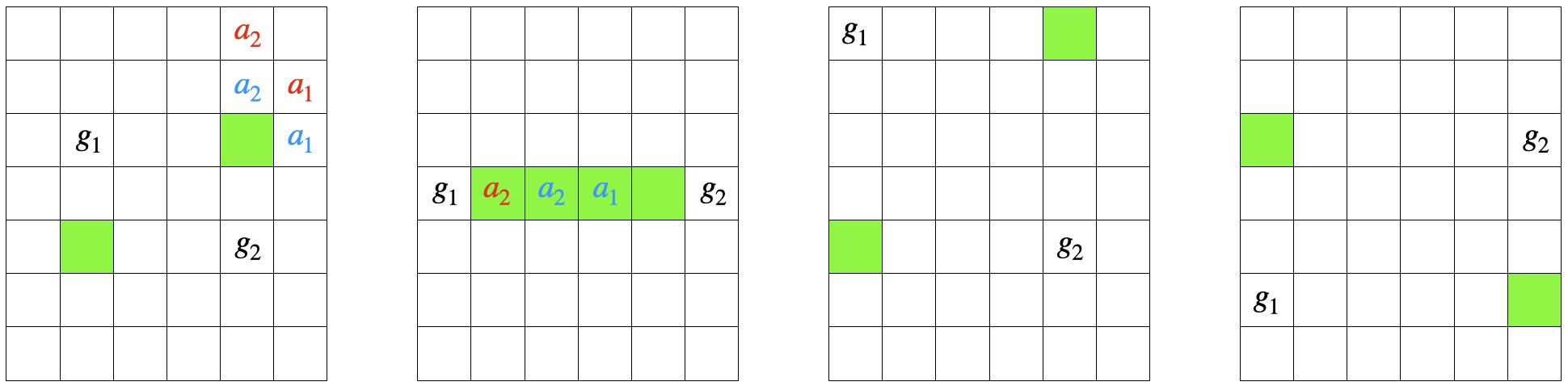}
\caption{(a) a typical case causing collisions; (b) a degraded case causing dead locks; (c) a corner case; (d) a boundary case}
\label{fig::accommodation}
\end{figure}

We then elaborate why \textit{crossroads} and \textit{streets} are crucial for feasible policies.
For example, Figure~\ref{fig::accommodation}(a) has shown that those blue agents $a_1$ and $a_2$ are running into an interior crossroad, and at the next step they will collide into the each other at this crossroad according to the rule of \textit{myopic preference}.
Note that for red ones in Figure~\ref{fig::accommodation}(a) can avoid collisions by both going down since it is indeed one of the most preferred actions, but then they will be faced with the same situation as the blue ones.
Similar situation happens in Figure~\ref{fig::accommodation}(b) for the blue $a_1$ and the red $a_2$ as they will collide into each other at the next step, while the blue $a_1$ and blue $a_2$ will simply run into a dead lock as they will both choose to temporally stop but will turn out to stop forever.
For Figure~\ref{fig::accommodation}(c) and (d), these goals do not form any interior crossroad and are not on the same street, and by experiments such configurations are solvable, which we will also prove.

We first prove a proposition about dead locks.
\begin{proposition}
	Given a map without obstacles with two agents, if a goal profile forms no interior crossroads, dead locks will only be formed iff two goals are on the same street, under the myopic preferences.
\label{prop::deadlock}
\end{proposition}

\begin{proof}\

\noindent
($\Leftarrow$) If two goals are on the same street, any two adjacent locations in between will cause a dead lock, given the \textit{myopic preference}.

\noindent
($\Rightarrow$) We prove the contraposition. If two goals are not on the same street,
there will not exist such a case when both agents have to stop, i.e., at least one agent has a better action moving itself closer to its goal.
\end{proof}

We can now prove Theorem \ref{thm::existence} as follows. It can be shown that such policy exists even with \textit{myopic preferences}, then it must also hold in general without this constraints.
\begin{proof}(of Theorem \ref{thm::existence}) For any of the two agents,
\begin{enumerate}
	\item If the only non-stop action that is the most preferred under the \textit{myopic preference} is blocked by the other adjacent agent, then it will stop to wait for exactly one time step, since there will no dead locks according to Proposition \ref{prop::deadlock}.
	\item If it is not the case of 1., there are at least two tied actions that are both the most preferred under the \textit{myopic preference}. Then there must be one of these two actions that will not cause any collision at the next step.
	\item Based on 1., 2. and Proposition \ref{prop::livelock}, we can see that for every time step, there is at least one agent shortening her distance towards her goal. Given that the map is finite, both agents will reach their goals in finite time steps. 
\end{enumerate}
\end{proof}

Moreover, we can claim that there is nothing to do with the shape of the map for Theorem \ref{thm::existence}
to hold.

\end{document}